\documentclass[12pt,superscriptaddress]{article}
\usepackage{palatino,cite,epsfig,amsmath,amssymb,color}

\def\nn {\nonumber}

\def\ti    {\tilde}

\def\nt   {{\ti \chi}^0}

\def\ctL  {{\cos2\Phi_L}}

\def\ctR   {\cos2\Phi_R}

\newcommand{\mnt}[1]   {m_{\ti \chi^0_{#1} }}
\newcommand{\mch}[1]   {m_{\ti \chi^\pm_{#1} }}

\newcommand{\lsim}{\;\raisebox{-0.9ex}{$\textstyle\stackrel{\textstyle<}
          {\sim}$}\;}

\newcommand{\eq}[1]{eq.~(\ref{#1})}
\newcommand{\fig}[1]{fig.~\ref{#1}}
\newcommand{\tab}[1]{table~\ref{#1}}
\definecolor{Black}{named}{Black}
\definecolor{Blue}{named}{Blue}
\definecolor{Red}{named}{Red}
\definecolor{Green}{named}{ForestGreen}
\definecolor{Black}{named}{Black}
\definecolor{Olive}{named}{OliveGreen}
\definecolor{Royal}{named}{RoyalBlue}
\definecolor{Orange}{named}{YellowOrange}
\definecolor{Yellow}{named}{Goldenrod}
\definecolor{Cornblue}{named}{CornflowerBlue}
\definecolor{Lila}{named}{DarkOrchid}

\begin{document}

\hfill hep-ph/0312069

\hfill IFT-03/34

\hfill IPPP-03-59

\hfill DCPT-03-118


\vspace{1cm}

\begin{center}

{\Large\bf SUSY Parameter Determination in Combined Analyses}

\vspace{0.3cm}

{\Large\bf at LHC/LC
}
 
\vspace{1cm}

{\sc 
K. Desch$^{1}$, J. Kalinowski$^{2}$, G. Moortgat-Pick$^{3}$,
M.M. Nojiri$^{4}$ \\ and  
G. Polesello$^{5}$
}

\vspace*{1cm}

{\sl
$^1$ Institut f\"ur Experimentalphysik, Universit\"at Hamburg,
  Notkestrasse 85, \\ D-22607 Hamburg, Germany

\vspace*{0.3cm}

$^2$Institute of Theoretical Physics, Warsaw University, Hoza 69,
\\ PL-00681 Warsaw, Poland

\vspace*{0.3cm}

$^3$IPPP, University of Durham, DH1 3LE Durham, UK

\vspace*{0.3cm}

$^4$YITP, Kyoto University, Kyoto 606 8502, Japan

\vspace*{0.3cm}

$^5$INFN, Sezione di Pavia,  Via Bassi 6, Pavia 27100, Italy

}

\end{center}

\vspace*{1cm}

\begin{abstract}
We demonstrate how the interplay of a future $e^+e^-$  LC at its first
stage with $\sqrt{s} \lsim 500$~GeV and of  the LHC 
could lead to a precise determination  of the
fundamental SUSY parameters in the gaugino/higgsino sector without
assuming a specific supersymmetry breaking scheme.  
We demonstrate this for  the benchmark
scenario SPS1a, taking into account 
realistic errors for the masses and cross sections
measured at the LC with polarised beams, including errors coming from
polarisation measurements, and mass measurements at
the LHC. The results 
clearly demonstrate the complementarity of the LHC and LC, and the
benefit from the joint analyses of their data.  

\end{abstract}


\section{Introduction}
Supersymmetry (SUSY) is one of the best motivated extensions of the
Standard Model (SM). However, since SUSY has to be broken even the 
minimal version, the unconstrained MSSM, has 105 new parameters. SUSY
analyses at future experiments, at the LHC and at a future Linear
Collider (LC), will have
to focus on the determination of these parameters in as
model-independent a way as possible \cite{ams03}. 

With so many new parameters clear strategies will be needed in
analysing the experimental data  \cite{parameters}. 
An interesting possibility 
to resolve the new physics is to start with the 
gaugino/higgsino particles which are expected to be among 
the lightest SUSY particles. 
At tree level, this sector depends
only on 4 parameters: $M_1$, $M_2$, $\mu$ and $\tan\beta$ -- the U(1)
and SU(2) gaugino masses, the higgsino mass parameter and the ratio of
the vacuum expectations of the two Higgs fields, respectively.   

Some strategies have been worked out for the determination at the tree
level the parameters $M_2$, $M_1$, $\mu$, $\tan\beta$ even if only
the light gaugino/higgsino particles, $\tilde{\chi}^0_1$,
$\tilde{\chi}^0_2$ and $\tilde{\chi}^\pm_1$ were kinematically
accessible at the first stage of the LC \cite{ckmz}.  In this report
we demonstrate how such an LC analysis could be strengthened if in
addition some information on the mass of the heaviest neutralino the LHC is
available.  We consider three scenarios:  (i) stand alone LC data, 
(ii) when the LC data are supplemented by the heavy neutralino
mass estimated from the LHC data, and (iii) joint analysis of the LC
and LHC  data. 
The results in the last scenario
will clearly demonstrate the essentiality of the LHC and LC and the
benefit from the joint analysis of their data.

In order to work out this hand-in-hand LHC+LC analysis for determining
the tree-level SUSY parameters, we assume that only the first
phase of a LC with a tunable energy up to $\sqrt{s}=500$~GeV 
would overlap with the LHC running. Furthermore, we assume an 
integrated luminosity of  $\int {\cal L} \sim 500$~fb$^{-1}$
and polarised beams with $P(e^-)=\pm 80\%$, $P(e^+)=\pm 60\%$. 
In the following $\sigma_L$ will refer to cross sections obtained
with $P(e^-)=- 80\%$, $P(e^+)= + 60\%$, and $\sigma_R$ with $P(e^-)=+
80\%$, $P(e^+)= - 60\%$. 
We restrict ourselves to 
the CP conserving chargino/neutralino sector and take the SPS1a as a
working benchmark \cite{sps}; the inclusion of 
CP violating phases will be considered elsewhere. 

Before presenting our results on the parameter determination, 
we first briefly recapitulate the main features of chargino and neutralino
sectors and sketch our strategy.

\section{ The gaugino/higgsino  sector}

\subsection{Chargino sector}
The mass matrix of the charged gaugino $\ti W^\pm$ and higgsino $\ti
H^\pm$   
is given by\footnote{One should note 
the difference between our convention of taking $\ti \chi^-$ 
as ``particles'' and e.g. the
convention of \cite{Haber-Kane}.}
\begin{eqnarray}
{\cal M}_C = \left(\begin{array}{cc}
     M_2              & \sqrt{2} m_W\cos\beta \\[2mm]
\sqrt{2} m_W \sin\beta & \mu\,
             \end{array}\right) \label{eq_charmat}
\end{eqnarray}
As a consequence of possible field redefinitions, the parameter
$M_2$ can be chosen
real and positive.
The two charginos $\tilde{\chi}^\pm_{1,2}$ are mixtures
of the charged SU(2) gauginos and higgsinos.
Since the mass matrix ${\cal M}_C$ is not symmetric,
two
different
unitary matrices acting on the left-- and right--chiral
$(\tilde{W},\tilde{H})_{L,R}$ two--component
states
\begin{eqnarray}
       \left(\begin{array}{c}
             \tilde{\chi}^-_1 \\
             \tilde{\chi}^-_2
             \end{array}\right)_{L,R} =
U_{L,R}\left(\begin{array}{c}
             \tilde{W}^- \\
             \tilde{H}^-
             \end{array}\right)_{L,R} 
\end{eqnarray}
define charginos as mass eigenstates. 
For real ${\cal M}_C$ 
the unitary matrices $U_L$ and $U_R$ can be parameterised as
\begin{eqnarray}
 U_{L,R}=\left(\begin{array}{cc}
             \cos\Phi_{L,R} & \sin\Phi_{L,R} \\
            -\sin\Phi_{L,R} & \cos\Phi_{L,R}
             \end{array}\right) 
\end{eqnarray}
The mass eigenvalues $m^2_{\tilde{\chi}^\pm_{1,2}}$ and the mixing
angles are given by
\begin{eqnarray}
 m^2_{\tilde{\chi}^\pm_{1,2}}
  & =&\frac{1}{2}(M^2_2+\mu^2+2m^2_W\mp \Delta_C)\nonumber \\
%
%
\cos 2\phi_{L,R}&=&-(M_2^2-\mu^2\mp 2m^2_W\cos 2\beta)/\Delta_C
   \nonumber
\end{eqnarray}
where 
$\Delta_C=[(M^2_2-\mu^2)^2+4m^4_W\cos^2 2\beta
              +4m^2_W(M^2_2+\mu^2)+8m^2_WM_2\mu
               \sin2\beta]^{1/2}$.

The $e^+e^-\to\tilde{\chi}^{\pm}_i \tilde{\chi}^{\mp}_j$ 
production processes occur
via the s-channel $\gamma$, $Z^0$ and the t-channel $\tilde{\nu}_e$
exchange.
Since the two mixing  angles $\Phi_{L,R}$ enter the couplings
in the $\ti \chi \ti\chi Z$ 
and  $e\ti\chi\ti\nu_e$ vertices, 
the chargino production cross sections 
$\sigma^\pm\{ij\}=\sigma(e^+e^-\to\tilde{\chi}^{\pm}_i
\tilde{\chi}^{\mp}_j)$ are  
bilinear functions of $\cos 2 \Phi_{L,R}$ \cite{CDGKSZ} and can be
written as 
\begin{equation}
\sigma^\pm\{ij\}=c_1 \cos^2 2 \Phi_L
+ c_2 \cos 2 \Phi_L + c_3  \cos^2 2 \Phi_R + c_4 \cos 2 \Phi_R
+ c_5 \cos 2 \Phi_L \cos 2 \Phi_R+c_6
\label{eq_sig11}
\end{equation}
We derived the coefficients $c_1,\ldots,c_6$
for the lightest chargino pair production cross section, see
\eq{eq_sig11} in the Appendix.

\subsection{Neutralino sector}
The neutralino mixing matrix in the 
$\{\tilde{\gamma}, \tilde{Z}^0, \tilde{H}^0_1, \tilde{H}^0_2\}$
basis is given by
\begin{eqnarray}
{\cal M}_N= \left(\begin{array}{cccc}
  M_1 \cos^2_W+M_2 \sin^2_W & (M_2-M_1) \sin_W \cos_W & 0  & 0 \\[2mm]
  (M_2-M_1) \sin_W \cos_W  & M_1 \sin^2_W+M_2 \cos^2_W &   m_Z  & 0\\[2mm]
0 & m_Z &       \mu \sin{2 \beta}       &     -\mu \cos{2 \beta} \\[2mm]
0 & 0 &     -\mu \cos{2 \beta}      &       -\mu \sin{2 \beta}
\end{array}\right)\
\label{eq_neutmat}
\end{eqnarray}
The neutralino eigenvectors and their masses 
are obtained with the 4$\times$4 diagonalisation
matrix $N$:
\begin{equation}
N {\cal M}_N N^{\dagger}={\sf diag}\{m_{\tilde{\chi}^0_1},\ldots,
m_{\tilde{\chi}^0_4}\}
\label{eq_neutev}
\end{equation}

The parameter $M_1$ can only be determined from the neutralino sector.
The characteristic equation of the mass matrix squared, 
${\cal M}_N {\cal M}^{\dagger}_N$, can be written
as a quadratic equation for the parameter $M_1$:
\begin{equation}
x_i M_1^2+y_i M_1-z_i=0,\quad\mbox{for}\quad i=1,2,3,4 \label{eq_m1}
\end{equation}
where $x_i$, $y_i$, $z_i$ are given by:
\begin{eqnarray}
x_i&=&- m_{\tilde{\chi}^0_i}^6+a_{41} m_{\tilde{\chi}^0_i}^4
-a_{21} m_{\tilde{\chi}^0_i}^2+a_{01},\label{eq_m1-1}\\
y_i&=&a_{42} m_{\tilde{\chi}^0_i}^4-a_{22} m_{\tilde{\chi}^0_i}^2+a_{02},
\label{eq_m1-2}
\\
z_i&=&m_{\tilde{\chi}^0_i}^8-a_{63} m_{\tilde{\chi}^0_i}^6
+a_{43} m_{\tilde{\chi}^0_i}^4-a_{23} m_{\tilde{\chi}^0_i}^2+a_{03},
\label{eq_m1-3}
\end{eqnarray}
The coefficients $a_{kl}$, ($k=0,2,4,6$, $l=1,2,3$), being    
invariants of the matrix ${\cal M}_N {\cal M}_N^T$, can be
expressed as  functions of $M_2$, $\mu$ and $\tan\beta$. Their explicit
form is given in the Appendix.

The $e^+e^-\to\tilde{\chi}^{0}_i \tilde{\chi}^{0}_j$ 
production processes occur
via the s-channel $Z^0$ and the t- and u-channel  $\tilde{e}_L$
and $\tilde{e}_R$ exchanges.
Since the neutralino mixing matrix $N$  is parameterised in general by
6 angles, the analytic expressions for the production cross sections
are more involved. Their explicit form can be found in \cite{ckmz}.

\begin{figure}[htb]
\setlength{\unitlength}{1cm}
\begin{center}
{\epsfig{file=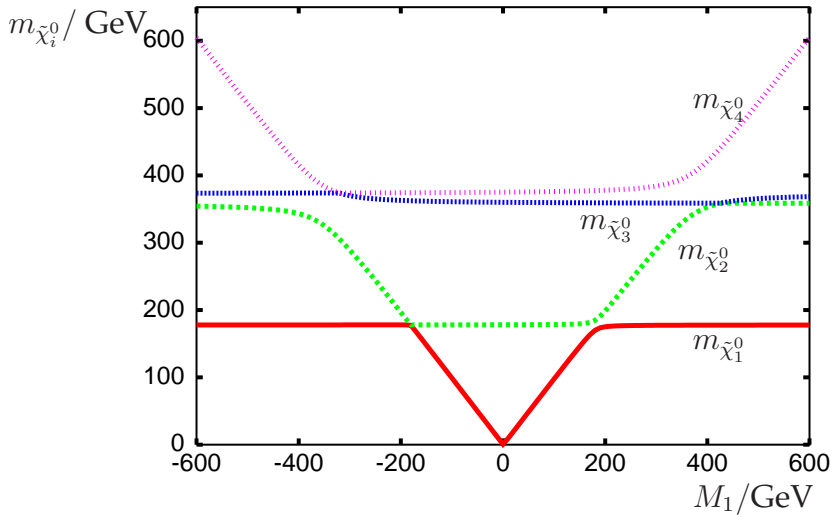,scale=.75}}
\put(-2,-.3){$M_1/$GeV}
\put(-11.1,6){$m_{\tilde{\chi}^0_i}/$~GeV}
\put(-2,1.8){$\mnt{1}$}
\put(-2.2,3){$\mnt{2}$}
\put(-3.5,3.4){$\mnt{3}$}
\put(-2,5){$\mnt{4}$}
\end{center}
\caption{\it $M_1$ dependence of the neutralino mass eigenvalues
$m_{\tilde{\chi}^0_i}$, $i=1,\ldots,4$ with $M_2$, $\mu$ and
$\tan\beta$ as in the reference scenario SPS1a. 
\label{fig_mass-ms}}
\end{figure}

As one can see from eq.~(\ref{eq_m1}) for each neutralino mass
$m_{\tilde{\chi}^0_i}$ one gets two solutions for $M_1$. 
In principle, a measurement
of two neutralino masses and/or the cross section resolves this
ambiguity.  
However, one has to remember that the mass eigenvalues
show different 
sensitivity to the parameter $M_1$, depending on their
gaugino/higgsino composition.  In
our scenario, the mass of the lightest neutralino $m_{\tilde{\chi}^0_1}$ 
depends
strongly on $M_1$ if $M_1$ is in the range $-183$~GeV$<M_1<180$~GeV, 
while the others are roughly insensitive, see
Fig.~\ref{fig_mass-ms}. For larger and larger $|M_1|$, the heavier
neutralinos become more sensitive to $M_1$ \cite{Moortgat-Pick:1999gp}. 

\subsection{The strategy}
At the initial phase of future $e^+e^-$ linear--collider operations
with polarised beams, the
collision energy may only be sufficient to reach the production 
thresholds of the light
chargino $\tilde{\chi}^\pm_1$ and the two lightest neutralinos
$\tilde{\chi}^0_1,\, \tilde{\chi}^0_2$.
From the analysis of this restricted system, nevertheless the entire 
tree level structure of the gaugino/higgsino sector can be unraveled
in analytical form 
in CP--invariant theories as follows~\cite{CDGKSZ,ckmz}.  

It is clear from eq.(\ref{eq_sig11}) that 
by analysing the $\tilde{\chi}^+_1\tilde{\chi}^-_1$ production cross
sections with polarised beams, $\sigma^\pm_L\{11\}$ and
$\sigma^\pm_R\{11\}$, 
the chargino mixing angles $\cos2\Phi_L$ and
$\cos2\Phi_R$ can be determined \cite{CDGKSZ}. 
Any two contours, $\sigma^\pm_L\{11\}$ and $\sigma^\pm_R\{11\}$ for example, 
will cross at least at one point in the plane between $-1 \leq \ctL,
\ctR \leq +1$, if the chargino and sneutrino masses are known
and the SUSY Yukawa coupling is identified with the gauge coupling.  
However, the contours, being of second order, 
may cross up to four times. 
The ambiguity can be resolved by measuring the transverse\footnote{The 
measurement of
the transverse cross section involves the azimuthal production angle
$\Phi$ of the charginos.  At very high energies their angle coincides
with the azimuthal angle of the chargino decay products. With
decreasing energy, however, the angles differ and the measurement of
the transverse cross section is diluted.} 
cross section $\sigma^\pm_T\{11\}$, or measuring $\sigma^\pm_L\{11\}$
and $\sigma^\pm_R\{11\}$ at different beam energies.

In the CP conserving case studied in this paper the SUSY parameters $M_2$,
$\mu$ and $\tan\beta$ can be determined  from the 
chargino mass $\mch{1}$ and mixing angles $\ctL$, 
$\ctR$~\cite{CDGKSZ}.  It is convenient to define
\begin{eqnarray}
p&=&\pm \left| \frac{\sin 2 \Phi_L+\sin 2 \Phi_R}{\cos 2 \Phi_L
-\cos 2 \Phi_R} \right| \label{eq_p}\\
q&=&\frac{1}{p} \frac{\cos 2 \Phi_L+\cos 2 \Phi_R}{\cos 2 \Phi_L
-\cos 2 \Phi_R} \label{eq_q}
\end{eqnarray}
Since the $\ctL$ and $\ctR$ are derived from 
$\tilde{\chi}^+_1 \tilde{\chi}^-_1$ cross sections,
the relative sign of $\sin 2\Phi_L$, $\sin 2 \Phi_R$ is not
determined and both possibilities in eqn.(\ref{eq_p}), (\ref{eq_q})
have to be considered.
From $p,q$, the SUSY parameters are 
determined as follows
($r^2=m^2_{\tilde{\chi}^{\pm}_1}/m^2_W$): 
\begin{eqnarray}
M_2&=&\frac{m_W}{\sqrt{2}}\left[(p+q)\sin\beta-
(p-q)\cos\beta\right] \label{eq_m2}\\
\mu&=&\frac{m_W}{\sqrt{2}}\left[(p-q)\sin\beta-(p+q)\cos\beta\right]
\label{eq_mu}\\
\tan\beta&=&
\left[\frac{p^2-q^2\pm \sqrt{r^2
(p^2+q^2+2-r^2)}} 
{(\sqrt{1+p^2}-\sqrt{1+q^2})^2- 2 r^2}\right]^\eta
\label{eq_tb1}
\end{eqnarray}
where $\eta=1$ for $\ctR > \ctL$, and $\eta=-1$
otherwise.  
The parameters $M_2$, $\mu$ are uniquely fixed if $\tan\beta$ is chosen 
properly. Since $\tan\beta$ is invariant
under simultaneous change of the signs of $p,q$, the definition 
$M_2 > 0$ can be exploited to remove this overall sign ambiguity.

The remaining parameter $M_1$ can be obtained from the neutralino
data \cite{ckmz}.  
The characteristic equation for the neutralino mass eigenvalues
\eq{eq_m1} is quadratic in $M_1$ if $M_2$, $\mu$ and $\tan\beta$
are already predetermined in the chargino sector.  In principle, two
neutralino masses are then sufficient to derive $M_1$. 
The cross
sections $\sigma_{L,R}^0\{12\}$ and $\sigma_{L,R}^0\{22\}$ for
production of $\nt_1\nt_2$ and $\nt_2\nt_2$ neutralino
pairs\footnote{The lightest neutralino--pair production cannot be
observed. Alternatively, one can try to exploit photon tagging in the
reaction $e^+e^- \rightarrow \gamma \tilde{\chi}^0_1\tilde{\chi}^0_1$
\cite{photon}.}  with polarised beams can serve as a consistency check
of the derived parameters.

In practice the above procedure may be much more involved due to
finite experimental errors of mass and cross section measurements,
uncertainties from  sneutrino and selectron masses which enter the cross
section expressions, errors on beam polarisation measurement, etc. In
addition, depending on the benchmark scenario, some physical
quantities in the light chargino/neutralino system may turn to be 
essentially
insensitive to some parameters. For example, as seen in
\fig{fig_mass-ms},  
the first two neutralino masses are insensitive to $M_1$ 
if $M_1 \gg M_2,\,\mu$. Additional information from the LHC
on heavy states, if available, can therefore be of great value in
constraining the SUSY parameters.

Our strategy can be applied only at the tree level. Radiative
corrections, which in the electroweak sector can be ${\cal O}$(10\%),  
inevitably bring all SUSY parameters
together \cite{radiative}. Nevertheless, tree level analyses should
provide in a  relatively 
model-independent way good estimates of SUSY parameters, which can be
further refined by including iteratively 
radiative corrections in an overall fit
to experimental data.  

\section{SUSY parameters from the LC data}

\subsection{ Experimental input  at the LC}

In this paper we adopt the SPS1a scenario defined at the electroweak
scale \cite{sps}. The  relevant SUSY parameters are 
\begin{eqnarray}
M_1=99.13~{\mbox GeV}, \quad M_2=192.7~{\mbox GeV}, \quad 
\mu=352.4~{\mbox GeV}, \quad \tan\beta=10
\end{eqnarray}
The resulting chargino and neutralino masses, together with the slepton
masses of the first generation, are given in \tab{tab_mass_LC}.

\begin{table}[h]
\begin{center}
\begin{tabular}{|c|cc|cccc|ccc|}
\hline
& ${\tilde{\chi}^{\pm}_1}$ & ${\tilde{\chi}^{\pm}_2}$ & 
${\tilde{\chi}^0_1}$ & ${\tilde{\chi}^0_2}$ &${\tilde{\chi}^0_3}$ &
${\tilde{\chi}^0_4}$ & ${\tilde{e}_R}$ &  ${\tilde{e}_L}$ &
${\tilde{\nu}_e}$ \\
\hline
mass & 
176.03 & 378.50 & 96.17 & 176.59 & 358.81 & 377.87 & 143.0 & 202.1 & 186.0 \\
error & 
0.55   &        & 0.05  & 1.2    &        &        & 0.05 & 0.2   & 0.7   \\
\hline 
\end{tabular}
\caption{\it Chargino, neutralino and slepton  masses in SPS1a, and 
the simulated experimental errors at the LC 
\cite{Martyn,Ball}. 
It is assumed that the heavy
chargino and neutralinos are not observed at the first phase of the
LC operating at $\sqrt{s}\le 500$~GeV. 
[All quantities are in GeV.]
\label{tab_mass_LC}}
\end{center} 
\end{table}

Because $\tilde{\chi}^+_1$ and $\tilde{\chi}^0_2$ decay dominantly
into $\tilde{\tau}$ producing the signal similar to that of stau pair
production, the $\tilde{\tau}$ mass and mixing angle are also
important for the study of chargino and neutralino sectors. The mass
and mixing angle can be determined as 
$m_{\tilde{\tau}_1}=133.2\pm 0.30$ 
GeV and
$\cos 2 \theta_{\tau}=-0.84\pm 0.04$, 
and the production cross section ranges from 43
fb to 138~fb depending on the beam polarisation, 
see \cite{stau,Martyn} for details of the stau
parameter measurements. We
assume that the contamination of stau production events can be
subtracted from the chargino and neutralino production. Below we
included the statistical error to our analysis but we did not include
the systematic errors.

\subsection{Chargino Sector}

As observables we use the light chargino mass and polarised cross
sections 
$\sigma^\pm_L\{11\}$ and $\sigma^\pm_R\{11\}$ at $\sqrt{s}=500$~GeV and 
$\sqrt{s}=400$~GeV. The light charginos $\ti \chi^\pm_1$ 
decay almost exclusively to
$\ti \tau^\pm_1 \nu_\tau$ followed by $\ti \tau^\pm_1 \to \tau^\pm \ti
\chi^0_1$. The signature for the $\ti{\chi}^\pm_1\ti{\chi}^\mp_1$
production would be two tau jets in opposite hemispheres
plus missing energy. 

The experimental errors that we assume and take into account
are:
\begin{itemize}
\item
The measurement of the chargino mass has  
been simulated and the expected error is 0.55~GeV,
\tab{tab_mass_LC}.
\item With $\int {\cal L}=500$~fb$^{-1}$
at the LC,  we assume 100~fb$^{-1}$ per each polarisation configuration
and we take into account 1$\sigma$ statistical error.
\item Since the chargino production is sensitive to $m_{\tilde{\nu}_e}$, 
we include its experimental error of 0.7~GeV.
\item The measurement of the beam polarisation with an uncertainty
of $\Delta P(e^{\pm})/P(e^ {\pm})=0.5\%$ is assumed. This error is  
conservative; discussions to reach errors smaller than  
$0.25\%$ are underway \cite{Power}.
\end{itemize}
The errors on production cross sections induced by the above
uncertainties, as well as the total errors (obtained by adding
individual errors in quadrature), are listed in 
\tab{tab_sig11}. 
We assume 100\% efficiency for the chargino cross sections 
due to a lack of realistic simulations.

\begin{table}
\begin{tabular}{|l|cc|cc|}
\hline
\phantom{++++}$\sqrt{s}$ &\multicolumn{2}{|c|}{400~GeV} &
\multicolumn{2}{|c|}{500~GeV} \\ 
($P(e^-)$, $P(e^+)$)
&$(-80\%,+60\%)$ &$(+80\%,-60\%)$ &
 $(-80\%,+60\%)$ & $(+80\%,-60\%)$\\ \hline
$\sigma(e^+e^-\to\tilde{\chi}^+_1\tilde{\chi}^-_1)$
& 215.84 & 6.38 & 504.87 &  15.07\\ \hline
$\delta\sigma_{\mbox{stat}}$ 
& 1.47 & 0.25 & 2.25 & 0.39\\ 
$\delta\sigma_{P(e^-)}$ &  
0.48 & 0.12 & 1.12 & 0.28 \\ 
$\delta\sigma_{P(e^+)}$ &
0.40 & 0.04 & 0.95 & 0.10 \\ 
$\delta\sigma_{m_{\tilde{\chi}^{\pm}_1}}$ &
7.09 & 0.20 & 4.27 & 0.12\\
$\delta\sigma_{m_{\tilde{\nu}_e}}$  & 
0.22 & 0.01 & 1.57 & 0.04 \\ \hline 
$\delta\sigma_{\mbox{total}}$ & 7.27 & 0.35 & 5.28 & 0.51 \\ \hline
\end{tabular}
\caption{\it Cross sections 
$\sigma_{L,R}^\pm\{11\}= 
\sigma_{L,R}(e^+ e^-\to \tilde{\chi}^+_1\tilde{\chi}^-_1)$ 
with polarised beams $P(e^-)=\mp 80\%$, $P(e^+)=\pm 60\%$
at $\sqrt{s}=400$ and 500~GeV and assumed errors (in fb) corresponding to 
 100~fb$^{-1}$ for each polarisation configuration. \label{tab_sig11}}
\end{table}

\begin{figure}[htb]
\setlength{\unitlength}{1cm}
\begin{center}
{\epsfig{file=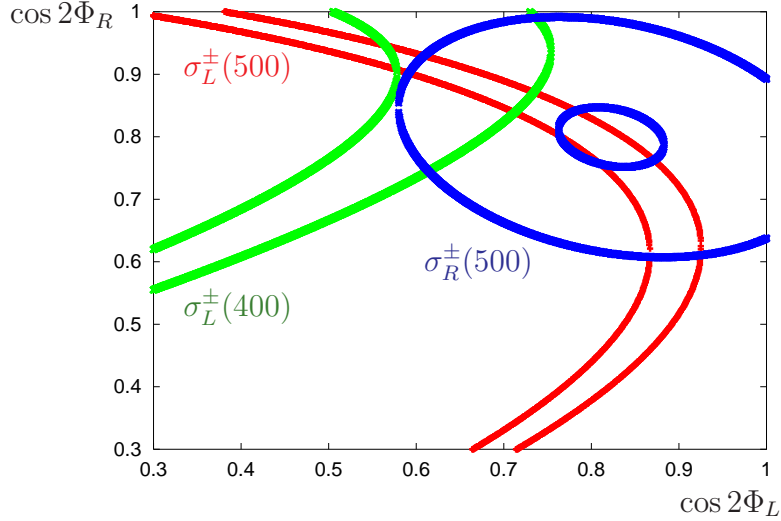,scale=.75}}
\put(-1.6,-.3){$\cos 2 \Phi_L$}
\put(-10.5,6.2){$\cos 2 \Phi_R$}
\put(-8.2,5.5){\color{Red}$\sigma_L^{\pm}(500)$}
\put(-8.2,2.3){\color{Olive}$\sigma_L^{\pm}(400)$}
\put(-5,2.9){\color{Blue}$\sigma_R^{\pm}(500)$}
\end{center}
\caption{\it $\cos 2 \Phi_R$ as a function of $\cos 2 \Phi_L$ 
for $\sigma^\pm_L\{11\}$ at $\sqrt{s}=500$~GeV (red),
and 400 GeV (green) and  $\sigma^\pm_R\{11\}$ at $\sqrt{s}=500$~GeV (blue)
within the error bounds (theo+exp) as given in \tab{tab_sig11}.
\label{fig_mix}}
\end{figure}

Now we can exploit the \eq{eq_sig11} and draw  
$\cos 2 \Phi_R
=f(\cos 2 \Phi_L,\sigma^\pm _{L,R}\{11\})$ 
consistent with the predicted cross sections
within the mentioned error bars, as  shown in \fig{fig_mix}. 
With the $\sqrt{s}=500$ GeV data alone two possible regions in the
plane are selected. With the help of the $\sigma^\pm_L\{11\}$ at
$\sqrt{s}=400$~GeV ($\sigma^\pm_R\{11\}$ is small and does not provide
further constraints) the
ambiguity is removed and  the mixing angles are limited  
within the range
\begin{eqnarray}
\cos 2 \Phi_L&=& [0.62,0.72] \label{eq_c2lrange}\\
\cos 2 \Phi_R&=& [0.87,0.91] \label{eq_c2rrange}
\end{eqnarray}

Although $\ctL,\ctR$ are determined rather precisely 
at a few per-cent accuracy, an
attempt to exploit eqns.~(\ref{eq_m2})-(\ref{eq_tb1}) shows that  
$M_2$ is reconstructed within 10 GeV, $\mu$ within 40 GeV, and
essentially no limit on $\tan\beta$ is obtained (we get
$\tan\beta>6$).
The main reason for this result is a relatively large error of  the
light chargino mass measurement due to the $\ti\chi^+_1\to \ti\chi^0_1
\tau^+\nu_\tau$ decay mode. Several methods exploiting other
sectors of the MSSM have been proposed to measure $\tan\beta$ in the
high $\tan\beta$ regime \cite{hightb,stau}. In the following we will 
exploit the neutralino sector (with 
eqns.~(\ref{eq_c2lrange}), (\ref{eq_c2rrange}) as the
allowed ranges for the chargino mixing angles) to improve constraints 
on $M_2$, $\mu$ and $\tan\beta$, and to determine $M_1$.  

\subsection{Neutralino Sector}

\begin{sloppypar}
As observables we use the two light neutralino masses and polarised
cross sections $\sigma^0_{L,R}\{12\}$ and $\sigma^0_{L,R}\{22\}$ at
$\sqrt{s}=400$~GeV and $\sqrt{s}=500$~GeV.  Although the production of
$\tilde{\chi}^0_1 \tilde{\chi}^0_3$ and $\tilde{\chi}^0_1
\tilde{\chi}^0_4$ pairs is kinematically accessible at
$\sqrt{s}=500$~GeV, the rates are small and the heavy states $\nt_3$
and $\nt_4$ decay via cascades to many particles.  Therefore we
constrain our analysis to the production of the light neutralino
pairs. 
\end{sloppypar}

\begin{table}[t]
\begin{tabular}{|l|cc|cc|}
\hline \phantom{++++} $\sqrt{s}$ 
&\multicolumn{2}{|c|}{400~ GeV} &
\multicolumn{2}{|c|}{500~GeV} \\ 
($P(e^-)$, $P(e^+)$)
&$(-80\%,+60\%)$ &$(+80\%,-60\%)$ &
 $(-80\%,+60\%)$ & $(+80\%,-60\%)$\\ \hline
$\sigma(e^+e^-\to \tilde{\chi}^0_1\tilde{\chi}^0_2)$
& 148.38 & 20.06 & 168.42 &  20.81\\ \hline
$\delta\sigma_{\mbox{stat}}$ 
& 2.92 & 1.55 & 3.47 & 1.55\\ 
$\delta\sigma_{\mbox{bg}}$
& 0.44 & 0.02 & 0.31 & 0.03\\ 
$\delta\sigma_{P(e^-)}$ &  
0.32 & 0.05 & 0.37 & 0.06 \\ 
$\delta\sigma_{P(e^+)}$ &
0.28 & 0.001 & 0.31 & 0.01 \\ 
$\delta\sigma_{m_{\tilde{\chi}^{\pm}_1}}$ &
0.21 & 0.30 & 0.16 & 0.26\\
$\delta\sigma_{m_{\tilde{e}_L}}$  & 
0.20 & 0.01 & 0.17 & 0.01 \\ 
$\delta\sigma_{m_{\tilde{e}_R}}$  & 
0.00 & 0.01 & 0.00 & 0.01 \\
\hline
$\delta\sigma_{\mbox{total}}$ & 3.0 & 1.58 & 3.52 & 1.57 \\ \hline
\end{tabular}
\caption{\it Cross sections 
$\sigma^0_{L,R}\{12\}=
\sigma_{L,R}(e^+ e^-\to \tilde{\chi}^0_1\tilde{\chi}^0_2)$
with polarised beams $P(e^-)=\mp 80\%$, $P(e^+)=\pm 60\%$
at $\sqrt{s}=400$ and 500~GeV,   and assumed 
errors  (in fb) corresponding to 
 100~fb$^{-1}$ for each polarisation configuration.\label{tab_sig12}}
\vspace{3mm}
\begin{tabular}{|l|cc|cc|}
\hline \phantom{++++} 
&\multicolumn{2}{|c|}{400~GeV} &
\multicolumn{2}{|c|}{500~GeV} \\ 
($P(e^-)$, $P(e^+)$)
&$(-80\%,+60\%)$ &$(+80\%,-60\%)$ &
 $(-80\%,+60\%)$ & $(+80\%,-60\%)$\\ \hline
$\sigma(e^+e^-\to\tilde{\chi}^0_2\tilde{\chi}^0_2)$
& 85.84 & 2.42 & 217.24 &  6.10\\ \hline
$\delta\sigma_{\mbox{stat}}$ 
& 2.4 & 0.4 & 3.8 & 0.6\\ 
$\delta\sigma_{P(e^-)}$ &  
0.19 & 0.05 & 0.48 & 0.12 \\ 
$\delta\sigma_{P(e^+)}$ &
0.16 & 0.02 & 0.41 & 0.05 \\ 
$\delta\sigma_{m_{\tilde{\chi}^{\pm}_1}}$ &
2.67 & 0.08 & 1.90 & 0.05\\
$\delta\sigma_{m_{\tilde{e}_L}}$  & 
0.15 & 0.004 & 0.28 & 0.01 \\ 
$\delta\sigma_{m_{\tilde{e}_R}}$  & 
0.00 & 0.00 & 0.00 & 0.00 \\
\hline
$\delta\sigma_{\mbox{total}}$ & 3.6 & 0.41 & 4.3 & 0.62 \\ \hline
\end{tabular}
\caption{\it Cross sections 
$\sigma^0_{L,R}\{22\}=
\sigma_{L,R}(e^+ e^-\to \tilde{\chi}^0_2\tilde{\chi}^0_2)$
with polarised beams $P(e^-)=\mp 80\%$, $P(e^+)=\pm 60\%$
at $\sqrt{s}=400$ and 500~GeV, and assumed 
errors (in fb) corresponding to 
 100~fb$^{-1}$ for each polarisation configuration. \label{tab_sig22}}
\end{table}

The neutralino $\ti{\chi}^0_2$ decays into $\ti{\tau}^\pm_1 \tau^\mp$
with almost 90\%, followed by the $\ti{\tau}^\pm_1 \to \tau^\pm
\ti{\chi}^0_1$. Therefore the final states for the
$\ti{\chi}^\pm_1\ti{\chi}^\mp_1$ and $\ti{\chi}^0_1\ti{\chi}^0_2$ are
the same (2$\tau$ + missing energy), however with different topology.
While for the charginos, the $\tau$'s tend be in opposite 
hemispheres with rather
large invariant mass, in the $\ti{\chi}^0_1\ti{\chi}^0_2$ process both
$\tau$'s, coming from the $ \ti{\chi}^0_2$ decay, would be more often 
in the same
hemisphere with smaller invariant mass. This feature allows 
separate the processes to some extent
exploiting e.g.~a cut on the opening angle
between the two jets of the  $\tau's$. However, in the case
of $\ti{\chi}^0_1\ti{\chi}^0_2$, significant background from
$\ti{\chi}^\pm_1\ti{\chi}^\mp_1$ and $\ti{\tau}^\pm_1 \ti{\tau}^\mp_1$
remains.

We estimate the statistical error on $\sigma(e^+e^-\to
\ti{\chi}^0_1\ti{\chi}^0_2)$ based the experimental simulation
presented in~\cite{Ball}. This simulation was performed at $\sqrt{s}
=500 $ GeV for unpolarised beams yielding an efficiency of 25\%. We 
extrapolate the statistical
errors at different $\sqrt{s}$ and different polarisations as
$\delta\sigma / \sigma = \sqrt{S+B}/S$ where we calculate the number
of signal (S) and background (B) events from the cross sections and
the integrated luminosity (100 fb$^{-1}$) assuming the same efficiency
as achieved for the unpolarised case. Since the cross sections for the
SUSY background processes are also known only with some uncertainty,
we account for this uncertainty in the background subtraction by
adding an additional systematic error ($\delta\sigma_{\mbox{bg}}$).

For the process $\tilde{\chi}^0_2 \tilde{\chi}^0_2 \to \tau^+\tau^-
\tau^+\tau^- \tilde{\chi}^0_1 \tilde{\chi}^0_1$
no detailed simulation exists. From the $\tau$-tagging efficiency
achieved in the $\ti{\chi}^0_1\ti{\chi}^0_2$ channel,
we assume that this final state can be reconstructed with an efficiency
of 15\% with negligible background. This is justified since no major
SUSY background is expected for the 4-$\tau$ final state,
$\mbox{BR}(\ti\nu_\tau\to\tau^+\tau^-\tilde{\chi}^0_1)^2$ is only  0.5\%.
SM backgrounds arise mainly from Z pair production and are small.

For both processes we account in addition for polarisation uncertainties
and uncertainties in the cross section predictions from the errors on
the chargino and selectron masses. Note that we implicitly assume that
the branching ratio $\ti{\chi}^0_2 \to \tau^+\tau^-\ti{\chi}^0_1$
is known, which is a simplifaction. A full analysis will have to
take into account the parameter dependence of this branching ratio
in addition, since it cannot be measured directly. 


The neutralino cross sections  depend on 
$M_1$, $M_2$, $\mu$, $\tan\beta$ and slepton masses. We prefer to
express $M_2$, $\mu$, $\tan\beta$ 
 in terms 
of $m_{\tilde{\chi}^{\pm}_1}$ and the mixing angles $\ctL,\ctR$. 
Then we consider  neutralino 
cross sections as  functions of unknown  
$M_1, \ctL,\ctR$ with uncertainties  due to statistics and 
experimental errors on beam polarisations,    
$m_{\tilde{\chi}^{\pm}_1}, m_{\tilde{e}_L}$ and  $m_{\tilde{e}_R}$ 
included (in quadrature) in the total error, see~\tab{tab_sig12}
and \tab{tab_sig22}.

\subsection{Results}
We perform a $\Delta \chi^2$ test defined as
\begin{equation}
\Delta \chi^2  =\sum_i |\frac{O_i -\bar O_i}{\delta O_i}|^2
\label{eq_chi2}
\end{equation}
The sum over physical observables $O_i$  includes
$m_{\nt_1},m_{\nt_2}$ and 
neutralino production cross sections
$ \sigma^0_{L,R}\{12\},\sigma^0_{L,R}\{22\}$ measured
at both energies of 400 and 500 GeV. 
The $\Delta \chi^2$ is a function of unknown $M_1,\ctL,\ctR$ with 
$\ctL,\ctR$ restricted to the ranges given in
eqns.~(\ref{eq_c2lrange}),(\ref{eq_c2rrange}) as 
predetermined from the chargino sector.
$\bar O_i$ stands for the physical observables 
taken at the input values of all
parameters, and $\delta O_i$ are the corresponding errors.

\begin{figure}[htb]
\setlength{\unitlength}{1cm}
\begin{center}
{\epsfig{file=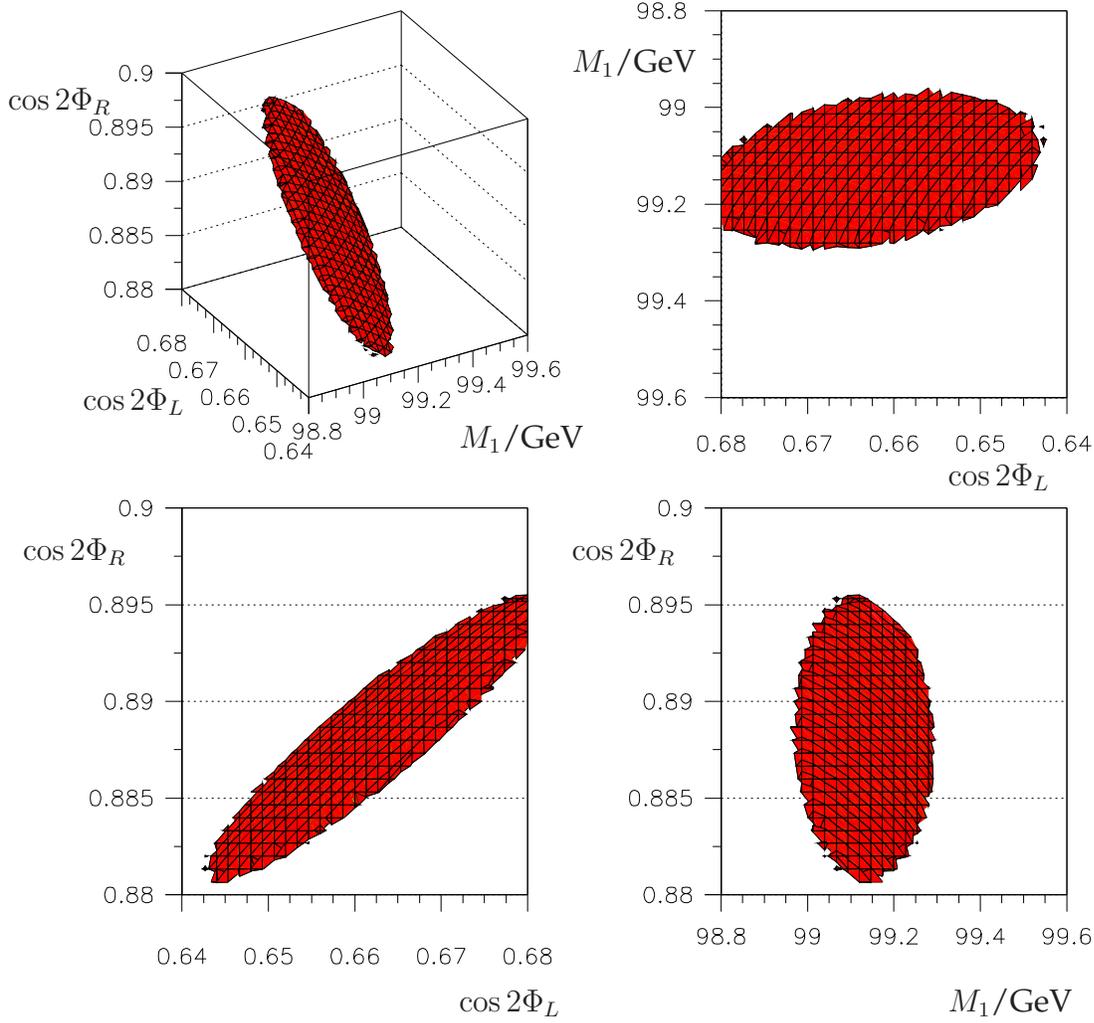,scale=.75}}
\put(-9.,0.){$\cos 2 \Phi_L$}
\put(-2.5,0.){$M_1$/GeV}
\put(-14.8,6.){$\cos2\Phi_R$}
\put(-7.5,6.){$\cos2\Phi_R$}
\put(-2.5,7.){$\cos 2 \Phi_L$}
\put(-7.5,12.5){$M_1$/GeV}
\put(-9.,7.5){$M_1$/GeV}
\put(-15.,12.){$\cos2\Phi_R$}
\put(-14,8.){$\cos 2 \Phi_L$}
\end{center}
\caption{\it 
The $\Delta \chi^2=1$ contour in the $M_1,\ctL,\ctR$ parameter space, and
its three 2dim projections, derived from the LC data.}
\label{fig_4cygLC}
\end{figure}

In \fig{fig_4cygLC} the contour of $\Delta \chi^2=1$ is shown in the
$M_1,\ctL,\ctR$ parameter space along with 
its three 2dim projections. The projection of the contours onto the
axes determines 1$\sigma$ errors for each parameter.

Values obtained for $M_1,\ctL,\ctR$ together with $\mch{1}$ can be
inverted to derive the fundamental parameters $M_2$, $\mu$ and
$\tan\beta$. At the same time masses of heavy chargino and neutralinos
are predicted. As can be seen
in \tab{tab_lc}, the parameters $M_1$ and $M_2$ are 
determined at the level of a few per-mil, while $\mu$ is reconstructed
within a few per-cent. Since the derived limits on $\tan\beta$ are
asymmetric, we show the interval  consistent with $\Delta\chi^2=1$.

\begin{figure}[h]
\begin{center}
\setlength{\unitlength}{0.8cm}
{\epsfig{file=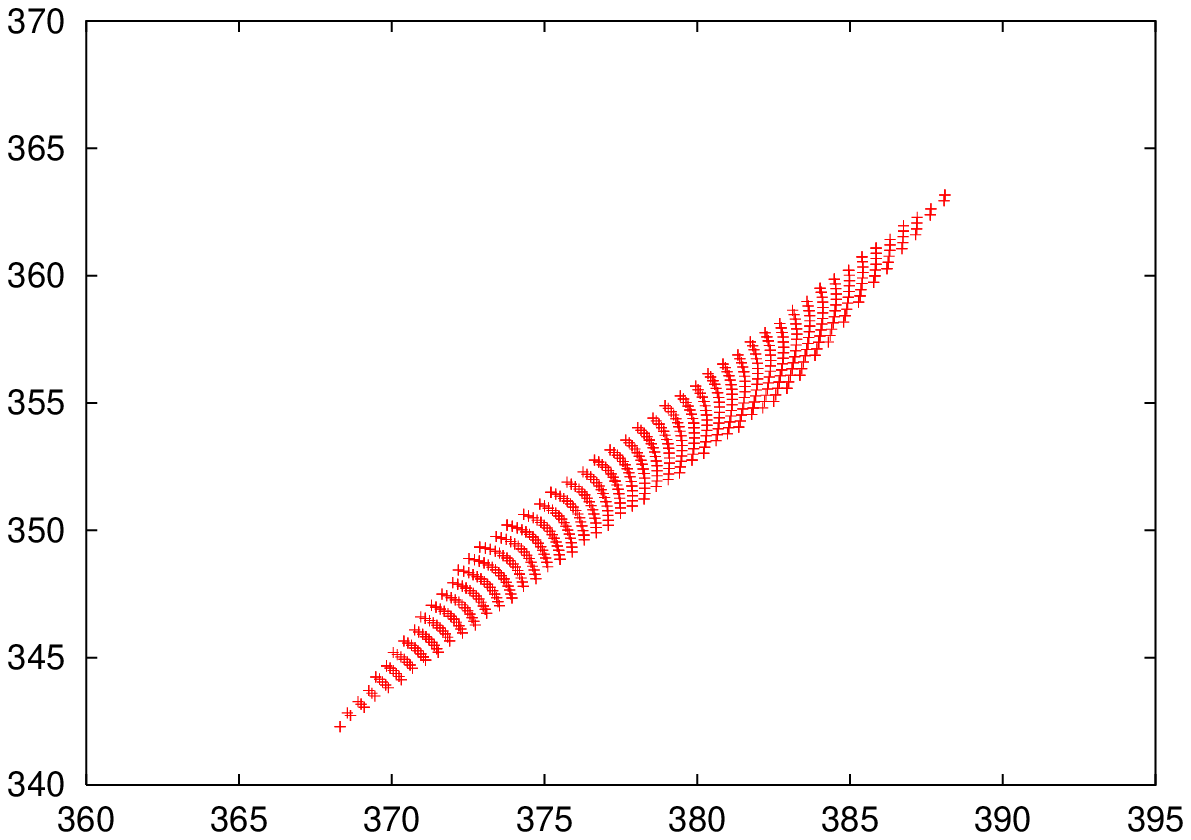,scale=.5}~~~~~~
  {\epsfig{file=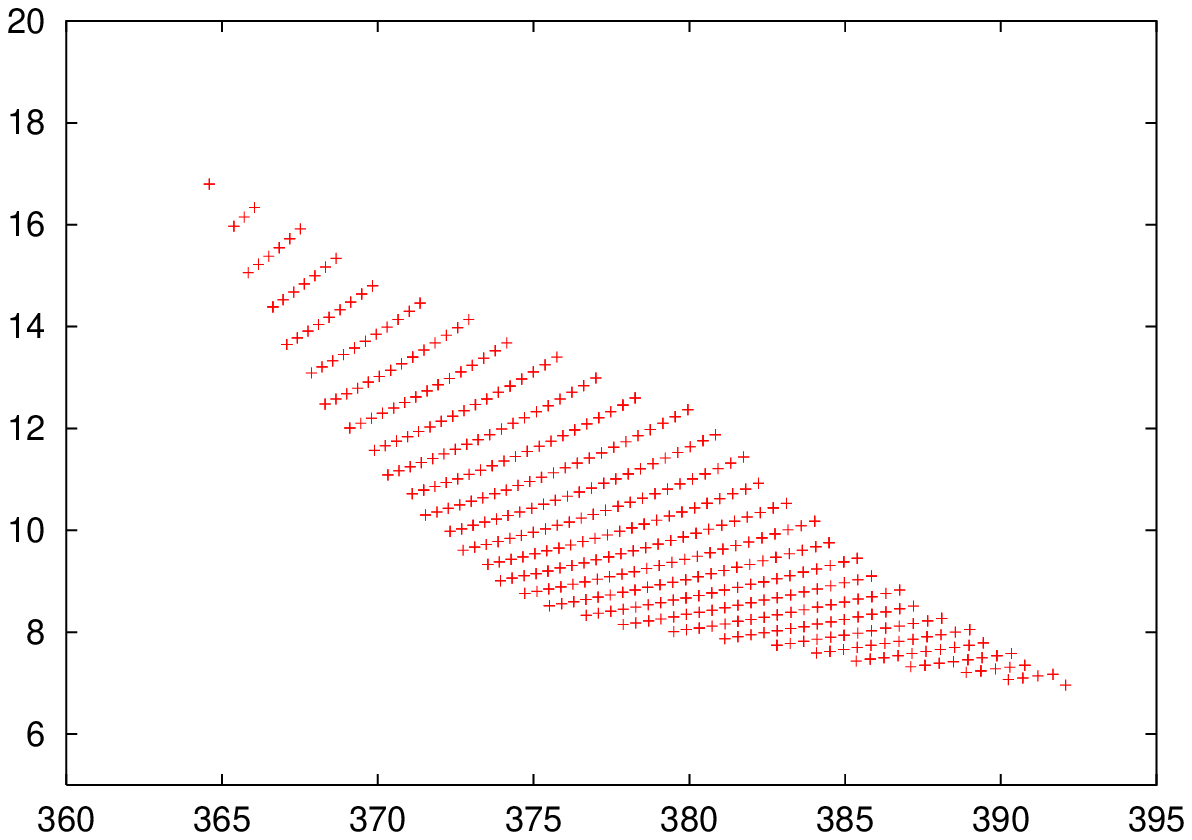,scale=.5}}}  
\put(-11.,-.3){\small $m_{\tilde{\chi}^0_4}/$GeV}
\put(-2,-.3){\small $m_{\tilde{\chi}^0_4}/$GeV}
\put(-18.3,5.){\small $\mu/$~GeV}
\put(-8.9,5.){\small $\tan\beta$}
\end{center}\par\vspace{-.5cm}
\caption{\it The correlation between predicted values of 
$\mu$ and $m_{\tilde{\chi}^0_4}$ (left panel) 
and the allowed range of 
$\tan\beta$ and 
$m_{\tilde{\chi}^0_4}$ (right panel)  from the analysis of the LC data.  
\label{fig-muchi04}}
\end{figure}

The errors on the predicted masses of the 
heavy chargino/neutralinos,  which in our
SPS1a scenario  are predominantly higgsinos, are strongly correlated
with the error of $\mu$;   the left panel of 
\fig{fig-muchi04} shows the correlation
between $\mu$ and $m_{\tilde{\chi}^0_4}$. In the right panel of this
figure a weaker correlation is
observed between $\tan\beta$ and $m_{\tilde{\chi}^0_4}$  (or between
$\tan\beta$ and  $\mu$).    Therefore,  by providing 
$m_{\tilde{\chi}^0_4}$ from endpoint measurements \cite{Giacomo},  
the LHC could considerably help to get a better
accuracy on $\mu$. At the same time a better determination of 
$\tan\beta$ can be expected.

\begin{table}
\begin{tabular}{|cccc|ccc|}
\hline
\multicolumn{4}{|c|}{SUSY Parameters}&
\multicolumn{3}{|c|}{Mass Predictions}\\
$M_1$ & $M_2$ & $\mu$ & $\tan\beta$ & $m_{\tilde{\chi}^{\pm}_2}$ &
$m_{\tilde{\chi}^0_3}$ & $m_{\tilde{\chi}^0_4}$ \\[2mm] \hline
$99.1\pm 0.2$ & $192.7\pm 0.6$ & $352.8\pm 8.9$ & $10.3\pm 1.5$
& $378.8 \pm 7.8$ & $359.2\pm 8.6$ & $378.2 \pm 8.1$\\ \hline
\end{tabular}
\caption{\it SUSY parameters with 1$\sigma$ errors 
derived from the  analysis of  the LC data
  collected at the first phase of operation. 
Shown are also the predictions for the heavier chargino/neutralino masses.
\label{tab_lc}}
\end{table}

\section{Combined strategy for the LHC and LC}

\subsection{LC data supplemented by ${\boldmath \mnt{4}}$  from the LHC}
The LHC experiments will be able to measure the masses of several
sparticles, as described in detail in \cite{Giacomo}.
In particular, the LHC will provide a first measurement of 
the masses of $\nt_1$, $\nt_2$ and $\nt_4$.
The measurements of $\nt_2$ and $\nt_4$ 
are achieved through the study of the processes:
\begin{equation}
\tilde\chi^0_i\rightarrow\tilde\ell\ell\rightarrow\ell\ell\nt_1
\end{equation}
where the index $i$ can be either 2 or 4.
The invariant mass of the two leptons in the final state shows 
an abrupt edge, which can be expressed in terms of the
masses of the relevant sparticles as
\begin{equation}
m_{l^+l^-}^{max} = m_{\nt_i} \sqrt{1-\frac{m^2_{\tilde\ell}}{m^2_{\nt_i}}}
\sqrt{1-\frac{m^2_{\nt_1}}{m^2_{\tilde\ell}}}
\label{eq_edge}
\end{equation}
If one only uses the LHC information, the achievable precision on
$m_{\nt_2}$ and $m_{\nt_4}$ will be respectively of 4.5 and 5.1 GeV
for an integrated luminosity of 300~fb$^{-1}$.\par 
In the case of the $\nt_4$, 
which  in the considered  scenario  is mainly higgsino, this information
can be exploited at the LC to constrain the parameter $\mu$  with a
better precision. 
If we include this improved precision on $\mnt{4}$ 
in the $\Delta\chi^2$ test of
\eq{eq_chi2}, the resulting $\Delta\chi^2=1 $ contours get modified  
and the achievable precision is improved, as shown in  \tab{tab_lc2}.

\begin{table}
\begin{tabular}{|cccc|cc|}
\hline
\multicolumn{4}{|c|}{SUSY Parameters}&
\multicolumn{2}{|c|}{Mass Predictions}\\
$M_1$ & $M_2$ & $\mu$ & $\tan\beta$ & $m_{\tilde{\chi}^{\pm}_2}$ &
$m_{\tilde{\chi}^0_3}$ \\ \hline
$99.1\pm 0.2$ & $192.7\pm 0.5$ & $352.4\pm 4.5$ & $ 10.2 \pm 0.9$
& $378.5 \pm 4.1$ & $358.8\pm 4.1$\\ \hline
\end{tabular}
\caption{\it SUSY parameters with 1$\sigma$ errors 
derived from the  analysis of  the LC data
  collected at the first phase of operation and
with   $\delta\mnt{4}=5.1$~GeV from the LHC.
Shown are also the predictions for the  masses of $\ti\chi^\pm_2$ and
  $\ti\chi^0_3$.
\label{tab_lc2}}
\end{table}

\subsection{Joint analysis of the LC
and LHC  data}

From the consideration of \eq{eq_edge}, one can see 
that the uncertainty on the LHC measurement of $\mnt{2}$ and $\mnt{4}$
depends both on the experimental error on the position of
$m_{l^+l^-}^{max}$, and on the uncertainty on $\mnt{1}$ and 
$m_{\tilde\ell}$. The latter uncertainty, which for both 
masses is of 4.8~GeV, turns out to be the dominant contribution.
A much higher precision can thus be achieved by inserting in 
\eq{eq_edge} the values for $\mnt{1}$, $m_{\tilde e_R}$  and 
$m_{\tilde e_L}$ which are measured at the LC with 
precisions respectively of 0.05, 0.05 and 0.2  GeV,  \tab{tab_mass_LC}. 

\begin{figure}[htb]
\setlength{\unitlength}{1cm}
\begin{center}
{\epsfig{file=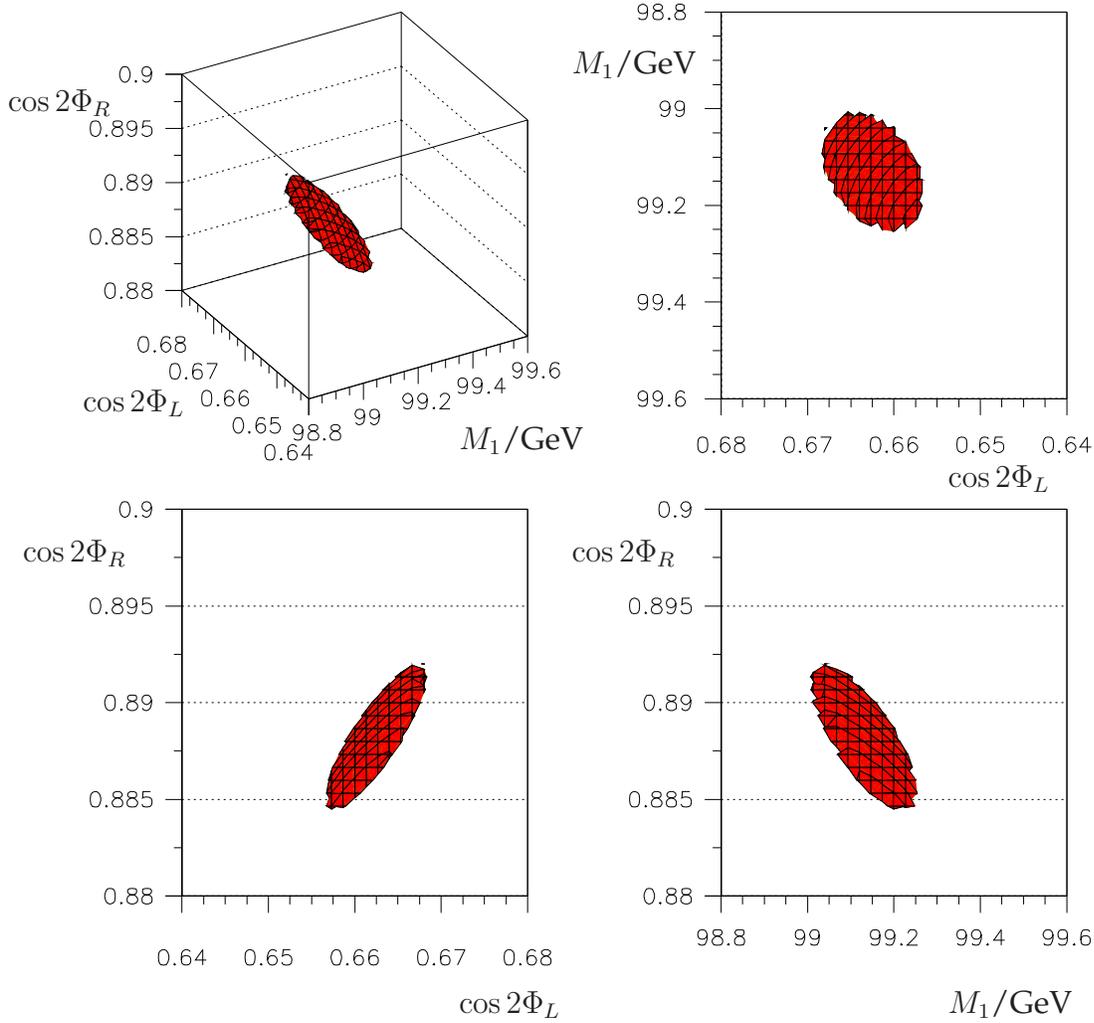,scale=.75}}
\put(-9.,0.){$\cos 2 \Phi_L$}
\put(-2.5,0.){$M_1$/GeV}
\put(-14.8,6.){$\cos2\Phi_R$}
\put(-7.5,6.){$\cos2\Phi_R$}
\put(-2.5,7.){$\cos 2 \Phi_L$}
\put(-7.5,12.5){$M_1$/GeV}
\put(-9.,7.5){$M_1$/GeV}
\put(-15.,12.){$\cos2\Phi_R$}
\put(-14,8.){$\cos 2 \Phi_L$}
\end{center}
\caption{\it The $\Delta \chi^2=1$ contour in the $M_1,\ctL,\ctR$ 
parameter space, and
its three 2dim projections, derived from the joint analysis of the LC
data and LHC data.}
\label{fig_4cygLHCLC}
\end{figure}

With this input the precisions on the LHC+LC measurements of 
$m_{\tilde{\chi}^0_2}$ and $m_{\tilde{\chi}^0_4}$ 
become: $\delta m_{\tilde{\chi}^0_2}=0.08$~GeV and  
$\delta m_{\tilde{\chi}^{0}_4}=2.23$~GeV.

\begin{table}[h]
\begin{tabular}{|cccc|cc|}\hline
\multicolumn{4}{|c|}{SUSY Parameters}&
\multicolumn{2}{|c|}{Mass Predictions}\\
$M_1$ & $M_2$ & $\mu$ & $\tan\beta$ & $m_{\tilde{\chi}^{\pm}_2}$ &
$m_{\tilde{\chi}^0_3}$ \\ \hline
$99.1\pm 0.1$ & $192.7\pm 0.3$ & $352.4\pm 2.1$ & $ 10.2 \pm 0.6$
& $378.5 \pm 2.0$ & $358.8\pm 2.1$\\ \hline
\end{tabular}
\caption{\it SUSY parameters  with 1$\sigma$ errors
derived from the combined analysis of the LHC and LC data
with $\delta m_{\tilde{\chi}^0_2}=0.08$~GeV and 
$\delta m_{\tilde{\chi}^0_4}=2.23$~GeV derived from the LHC when using 
the LC input of $\delta m_{\tilde{\chi}^0_1}=0.05$~GeV.
\label{tab_lhclc}}
\end{table}

From the results of the $\Delta\chi^2$ test one can calculate  the 
improvement in accuracy 
for the derived parameters by imposing the new mass constraints. The
final results are shown in 
\tab{tab_lhclc}.  The accuracy for the parameters $\mu$ and
particularly $\tan\beta$ is much better, as could be
expected from  
\fig{fig-muchi04}, where the
allowed range of $\mu$ and $\tan\beta$ 
from the LC analysis is  considerably reduced once 
the measured mass $m_{\tilde{\chi}^0_4}$ at the LHC
is taken into account. In particular, the precision on $\tan\beta$ 
becomes better than from  other SUSY sectors
\cite{hightb,stau}.

\section{Summary }
We have worked out in a specific example, a mSUGRA scenario with rather high
$\tan\beta=10$, how the combination of the results from 
the two accelerators, LHC and LC,  allows
a precise determination of  the fundamental 
SUSY parameters without assuming a specific supersymmetry breaking scheme.  

Measuring with high precision
the masses of the expected lightest SUSY particles $\tilde{\chi}^0_1$,
$\tilde{\chi}^0_2$, $\tilde{\chi}^{\pm}_1$ and their cross sections at 
the LC, and 
taking into account simulated mass measurement errors and
corresponding uncertainties for the theoretical predictions, we could
determine the fundamental SUSY parameters $M_1$, $M_2$, $\mu$ 
at tree level within a
few percent, while  $\tan\beta$ is estimated within $\sim$ 20 \%.
The masses of  heavier chargino and neutralinos can also be predicted
at a level of a few percent. The use of polarised beams at the  LC
is decisive for deriving unique solutions.

If the LC analysis is supplemented with the LHC measurement of the
heavy neutralino mass, the errors on $\mu$ and $\tan\beta$ can be
improved. However, the best results are obtained when first  
the LSP and slepton masses from the LC
are fed to the LHC analyses to get a precise experimental determination 
of the $\nt_2$ and $\nt_4$ masses, 
which  in turn are  injected back to the analysis of the
chargino/neutralino LC data. Such a combined strategy 
will provide  in particular the precise measurement of the 
$\nt_4$ mass, the  parameters $\mu$  
with an accuracy at the $\le O(1\%)$ level, and 
the error for $\tan\beta$ of the order of $\le$ 10\%,  reaching a
stage where radiative corrections become relevant in the electroweak
sector and which will have to be taken into account in  future fits 
 \cite{radiative}.

\subsection*{Acknowledgments}
The authors would like to thank G. Weiglein for motivating this
project and for the coordination of the LHC/LC working group.  
We are indebted to 
G. Blair, U. Martyn  and W. Porod for useful
discussions and for providing the errors for simulated mass 
and cross section measurements at the LC. 
The work is supported in part by the European Commission 5-th
Framework Contract HPRN-CT-2000-00149.
JK was supported by the KBN Grant 
2 P03B 040 24 (2003-2005).

\section*{Appendix}
{\bf a)} For the lightest chargino pair production, $\sigma^\pm\{11\} =\sigma(
e^-(p_1) e^+(p_2)\to \tilde{\chi}^+_1(p_3)\tilde{\chi}^-_1(p_4))$, 
the coefficients 
$c_1,\ldots,c_6$ in \eq{eq_sig11} are given by:
\begin{eqnarray}
c_1&=& \int_C |Z|^2 
  \{c_{LR} L^2 f_2+c_{RL} R^2 f_1\} \nn 
\\
c_2&=&\int_C |Z|^2 \{c_{LR} L^2 (1-4L)(2 f_2+f_3)+c_{RL} R^2(1-4R)
(2f_1+f_3)\}\nonumber\\
&&-\int_C G\tilde{N} 4
\{c_{LR} L (2 f_2+f_3)+c_{RL}R(2 f_1+f_3)\}
-\int_C Re(Z)\tilde{N} c_{LR} L f_3
\nn 
\\
c_3&=& \int_C |Z|^2 (c_{LR} L^2 f_1+c_{RL} R^2 f_2)
-\int_C Z\tilde{N} 2 c_{LR} L f_1
+\int_C \tilde{N}^2 c_{LR} f_1 \nn 
\\
c_4&=&\int_C  |Z|^2 (1-4L)\{ c_{LR} L^2 (2 f_1+f_3)
+c_{RL} R^2 (2 f_2+f_3)\}+\int_C \tilde{N}^2 2 c_{LR} f_1
\nonumber\\
&&
+\int_C Re(Z)\tilde{N} c_{LR} L \{-4f_1 -f_3+4L(2f_1+f_3)\}
+\int_C G\tilde{N} 4 c_{LR} (2 f_1+f_3)
\nonumber \\
&&
-\int_C G Re(Z) 4 \{c_{LR} L (2 f_1+f_3)
+c_{RL} R (2 f_2+f_3)\} \nn 
\\
c_5&=& \int_C |Z|^2 (c_{LR} L^2+c_{RL} R^2)f_3
-\int_C Re(Z)\tilde{N} c_{LR} L f_3 \nn 
\\
c_6&=&\int_C  |Z|^2 \{c_{LR} L^2 (1-8L)+c_{RL} R^2 (1-8L)
+16 L^2 (c_{LR} L^2+c_{RL} R^2)\}(f_1+f_2+f_3)
\nonumber\\
&&-\int_C Re(Z)\tilde{N} c_{LR} L (1-4L) (2 f_1+f_3)
+\int_C G^2 (c_{LR}+c_{RL}) (f_1+f_2+f_3)
\nonumber\\
&& -\int_C Re(Z)G 8 \{c_{RL} R + c_{LR} L (1-4L)\}
(f_1+f_2+f_3)
+\int_C \tilde{N}^2 c_{LR} f_1
 \nonumber\\
&&
+\int_C G \tilde{N} 4 c_{LR} (2 f_1+f_3) \nn
\end{eqnarray}
where 
$\int_C=\frac{q_{\tilde{\chi}}}{E_b^3}\frac{1}{2 \pi}\int {\rm d}\cos\theta$, 
$L=-\frac{1}{2}+\sin^2\theta_W$, $R=\sin^2\theta_W$, and  
\begin{eqnarray*}
G&=&e^2/s,\quad 
Z=g^2/\cos^2\theta_W(s-m_Z^2+i m_Z \Gamma_Z),\quad
\tilde{N}=g^2/(t-m_{\tilde{\nu}_e}^2) \nn 
\end{eqnarray*}
denote the $\gamma$, $Z$ and $\ti\nu_e$ propagators,  
\begin{eqnarray*}
c_{LR}&=&(1-P(e^-))(1+P(e^+)),\quad
c_{RL}=(1+P(e^-))(1-P(e^+))
\end{eqnarray*}
are the beam polarisation factors, and  
\begin{eqnarray*}
f_1=(p_1 p_4)(p_2 p_3), \quad 
f_2=(p_1 p_3)(p_2 p_4), \quad 
f_3=s \, m^2_{\tilde{\chi}^\pm_i}  /2
\end{eqnarray*}
are the pure kinematic
coefficients.

{\bf b)} The coefficients 
$a_{kl}$ ($k=0,2,4,6$, $l=1,2,3$), which appear in
eqns.~(\ref{eq_m1-1}),(\ref{eq_m1-2}) and (\ref{eq_m1-3}), are   
invariants of the matrix ${\cal M}_N {\cal M}_N^T$. They can be
expressed as  functions of
the parameters 
$M_2$, $\mu$, $\tan\beta$ 
in the  following way:
\begin{eqnarray*}
a_{63}&=&M_2^2+2 (\mu^2+m_Z^2)\label{eq_inv1}\\
a_{41}&=&M_2^2+2 (\mu^2+m_Z^2 \cos^2\theta_W) \label{eq_inv2}\\
a_{42}&=&-2 \mu m_Z^2 \sin2 \beta \sin^2\theta_W\label{eq_inv3}\\
a_{43}&=&2 \mu^2 M_2^2+(\mu^2+m_Z^2)^2-2 m_Z^2\mu M_2\sin2 \beta\cos^2\theta_W
+2 m_Z^2 M_2^2 \sin^2\theta_W \label{eq_inv4}\\
a_{21}&=& \mu^4+2 \mu^2 M_2^2+2 m_Z^2 \mu^2 \cos^2\theta_W
+m_Z^4 \cos^2\theta_W-2 m_Z^2 M_2 \mu \sin2 \beta \cos^2\theta_W
\label{eq_inv5}\\
a_{22} &=&2 [ m_Z^4 M_2 \sin^2\theta_W \cos^2\theta_W
- m_Z^2 \mu^3 \sin^2\theta_W \sin2 \beta
-m_Z^2 \mu M_2^2 \sin^2\theta_W \sin2\beta] \label{eq_inv6}\\
a_{23}&=&\mu^4 M_2^2+m_Z^4 \mu^2 \sin^2 2\beta
+2 m_Z^2 \mu^2 M_2^2 \sin^2\theta_W
-2 m_Z^2 M_2 \mu^3\cos^2\theta_W\sin2\beta+m_Z^4 M_2^2\sin^4\theta_W
\nonumber
\label{eq_inv7}\\
a_{01}&=& \mu^4 M_2^2+m_Z^4 \mu^2 \cos^4\theta_W\sin^2 2\beta-2 m_Z^2\mu^3 M_2
\cos^2\theta_W \sin2 \beta\label{eq_inv8}\\
a_{02}&=&2 m_Z^4 \mu^2 M_2 \sin^2\theta_W \cos^2\theta_W \sin^2 2 \beta
-2 m_Z^2 \mu^3 M_2^2\sin^2\theta_W \sin2 \beta \label{eq_inv9}\\
a_{03}&=& m_Z^4 \mu^2 M_2^2 \sin^4\theta_W \sin^2 2 \beta
\end{eqnarray*}



\end{document}